\documentclass[a4paper,
               ]{jacow}
\usepackage{pdfpages,multirow,ragged2e} %
\makeatletter%
	\ifboolexpr{bool{xetex}}
	 {\renewcommand{\Gin@extensions}{.pdf,%
	                    .png,.jpg,.bmp,.pict,.tif,.psd,.mac,.sga,.tga,.gif,%
	                    .eps,.ps,%
	                    }}{}
\makeatother

\ifboolexpr{bool{xetex} or bool{luatex}} 
 {}                                      
 {\usepackage[utf8]{inputenc}}           

\usepackage[USenglish]{babel}

\ifboolexpr{bool{jacowbiblatex}}%
 {%
  \addbibresource{jacow-test.bib}
  \addbibresource{biblatex-examples.bib}
 }{}
\begin{document}

\title{Updated baseline design for HALHF: \\ the hybrid, asymmetric, linear Higgs factory}

\author{C. A. Lindstr{\o}m\thanks{c.a.lindstrom@fys.uio.no}, E. Adli,  J. B. B. Chen, P. Drobniak, E. E. Hørlyk, D. Kalvik, K. N. Sjobak, \\ Department of Physics, University of Oslo., Oslo, Norway \\
T. Barklow, S. Gessner, M. Hogan, SLAC National Accelerator Laboratory, Menlo Park, CA, USA \\
M. Berggren, A. Laudrain, B. List, J. List, V. Maslov, K. P{\~o}der, M. Th{\'e}venet, N. Walker, J. Wood, \\ Deutsches Elektronen-Synchrotron DESY, Hamburg, Germany \\
S. Boogert, Cockcroft Institute, Daresbury Laboratory, STFC, Warrington, UK \\
P. N. Burrows, V. Cilento, R. D'Arcy, B. Foster, \\ John Adams Institute, Department of Physics, University of Oxford, Oxford, UK \\
S. Farrington, Rutherford Appleton Laboratory, STFC, Didcot, UK \\
 X. Lu, Northern Illinois University, USA \\
 G. Moortgat-Pick, University of Hamburg, Germany \\
 A. Seryi, Thomas Jefferson National Accelerator Facility, Newport News, VA, USA
 }
 
\maketitle

\begin{abstract}
    Particle physicists aim to construct a electron--positron Higgs factory as the next major particle collider. However, the high associated costs motivate the development of more affordable collider designs. Plasma-wakefield acceleration is a promising technology to this end. HALHF is a proposal for a Higgs factory that utilizes beam-driven plasma-wakefield acceleration to accelerate electrons to high energy with high gradient, while using radio-frequency acceleration to accelerate positrons to a lower energy. This asymmetry sidesteps a major difficulty in plasma acceleration: that of accelerating positrons with high efficiency and quality. Since publication, several challenges were identified in the original baseline design. We summarize the updated baseline design, which addresses these challenges, and describe the parameter- and cost-optimization process used to arrive at this design.
\end{abstract}

\section{Introduction}

The 2020 update of the European Strategy for Particle Physics \cite{ESPP2020} and the 2023 report from the Particle Physics Project Prioritization Panel (P5) \cite{P5} both identify a Higgs factory as the next flagship collider for particle physics. Several mature proposals exist \cite{CLIC,ILC,FCC}, but the cost estimates approach or exceed 10~BCHF. This cost issue has motivated a push to develop more affordable accelerator technology. Plasma acceleration~\cite{Veksler1956,Fainberg1956,Tajima1979,Chen1985,Ruth1985}, which uses high-intensity laser \cite{Esarey2009} or particle beams \cite{LindstromCorde2025} to drive strong accelerating fields in a plasma, is one of the most promising directions. Several plasma-based collider concepts have been proposed \cite{Seryi2009,Adli2013}, each addressing challenges raised in the previous one, gradually becoming more concrete and self-consistent.

A recently proposed plasma-based collider is HALHF (a hybrid, asymmetric, linear Higgs factory) \cite{Foster2023}, which builds on previous proposals---e.g., in that it uses electron beam drivers for high power-efficiency---but addresses a major unsolved problem: positron acceleration. Currently, it is not clear how to utilize plasmas, which are charge asymmetric due to the mass imbalance of electrons and ions, to accelerate positrons with the same energy efficiency and beam quality as is possible with electrons \cite{Cao2024}. HALHF therefore accelerates only the colliding electrons in a multistage plasma accelerator \cite{Lindstrom2021a} while accelerating positrons using a more conventional radio-frequency (RF) accelerator. To minimize the overall size and cost of the facility, the energy of the electrons, accelerated compactly, must be significantly higher than that of the positrons. It was also found that asymmetric bunch charges (i.e., more positrons) are beneficial for overall power efficiency, and that asymmetric emittances (i.e., higher-emittance electrons) are both possible and desirable for relaxed tolerances in the plasma arm.

Over the past two years, the HALHF Collaboration has studied the implications of such an hybrid and asymmetric design. While the core concept still appears viable, several aspects have been identified as either problematic or challenging, including the use of a combined-function RF linac for both colliding positrons and driver electrons, use of tight high-energy bends, high positron charges and the required cooling rate of plasma cells. An updated baseline design was therefore recently proposed \cite{Foster2025}, with changes made to both the overall geometry as well as detailed parameters.

\begin{figure*}[t]
	\centering
    \includegraphics[width=\textwidth]{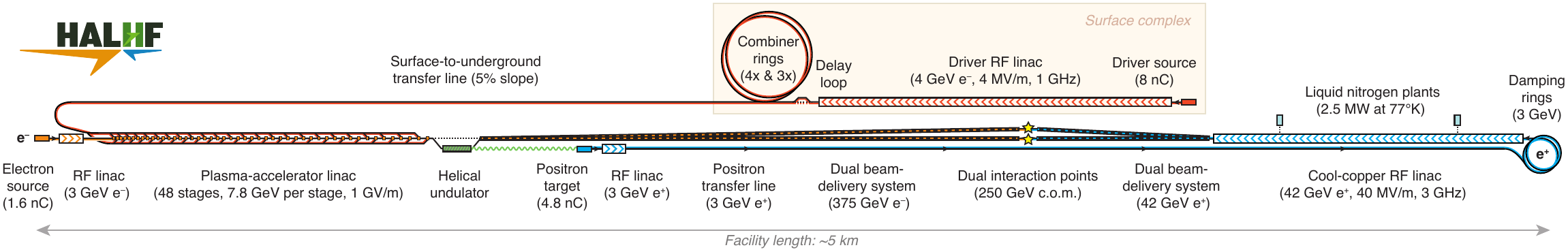}
\caption{Schematic view of the 250-GeV version of the updated baseline design for HALHF. From Ref.~\cite{Foster2025}.}
    \label{fig:new-baseline}
\end{figure*}

In this paper, we summarize the changes in the updated baseline, HALHF 2.0 (see Fig.~\ref{fig:new-baseline}), as reported in greater detail in Refs.~\cite{Foster2025,HALHF2025,HALHF2025backup}, and focus in particular on describing the optimization process used to arrive at this parameter set.

\section{Major changes from the \\ original baseline design}

The largest change from the original baseline is the use of separate positron and driver linacs: this allows the gradient and time structure of each linac to be separately optimized for positrons (i.e., high gradient and low current) and driver electrons (i.e., low gradient and high current). This approach also avoids the need to develop a novel combined-function linac, instead leveraging existing cost-optimized designs. The proposal for the drive-beam linac is to use a CLIC-like RF drive-beam linac (frequency \SI{1}{GHz} and gradient \SI{4}{MV/m}), accelerating \SI{8}{nC} bunches with high wall-plug-to-beam power efficiency ($\sim$50\%). It uses a delay loop and two combiner rings that frequency multiply a long, low-current bunch train (\SI{4}{ns} spacing) to a shorter, higher-current bunch train (\SI{0.17}{ns} spacing), as this reduces the peak RF power and is more easily distributed to the multistage plasma accelerator. The positron RF linac is based on cool-copper technology, cooled to liquid-nitrogen temperatures, and operated at frequency \SI{3}{GHz} and gradient \SI{40}{MV/m} (conservative compared to C$^3$ \cite{Vernieri2023}). The linac separation also allows use of more numerous but lower-energy drivers: 48 stages (instead of 16 in the original proposal) and a driver energy of \SI{4}{GeV} (instead of \SI{31}{GeV}), significantly easing the transport and distribution of these drivers.

The accelerating gradient in the plasma has been reduced from \SI{6.4}{GV/m} to \SI{1}{GV/m}, with a corresponding reduction in plasma density from \SI{7e15}{\per\cubic\cm} to \SI{6e14}{\per\cubic\cm}. This is motivated by the observations that the plasma linac is not a significant cost driver unless the gradient is below \SI{\sim 1}{GV/m} and that several issues in the original design are mitigated at reduced density---such as synchronization and misalignment tolerances, matching, beam ionization and plasma cooling.

Another change, motivated by physics performance, is the introduction of a polarized positron source, modeled on the helical undulator used for ILC \cite{Baynham2009}. The interaction point (IP) region is upgraded to feature two detectors, and consequently a dual beam-delivery system---this is envisioned to follow the dual-IP scheme proposed by CLIC \cite{Cilento2021}.

Finally, the overall energy asymmetry between electrons and positrons is reduced in HALHF 2.0; from \SI{500}{GeV} versus  \SI{31}{GeV}, respectively (for the \SI{250}{GeV} center-of-mass design), to \SI{375}{GeV} versus \SI{41}{GeV}. This change minimizes the overall collider length (taking into account both arms) and the overall construction cost when all the associated changes are taken into account, as described below.

\section{Defining an optimization metric}

Choosing an optimization metric for a collider is non-trivial. Simple metrics like length, power and luminosity are incomplete, as minimizing one individually explodes others. Cost is a more meaningful metric, as it allows several metrics such as length and power to be balanced. One way to balance these is by defining a ``Full Programme Cost'', a complex metric that encompasses the entire cost of building and operating the collider until the desired amount of data has been gathered (e.g., \SI{2}{ab^{-1}}):
\begin{align*}
    \mathrm{Full~Programme~Cost}=&\mathrm{~} \mathrm{Construction~Cost} + \mathrm{Overheads} \\&\mathrm{~}+ \mathrm{Integrated~Energy~Cost} \\&\mathrm{~}+ \mathrm{Maintenance~Cost} \\&\mathrm{~}+ \mathrm{Carbon~Shadow~Cost}
\end{align*}
The construction cost includes both the machine components and all the civil engineering (tunnels, buildings, IP), including infrastructure and services such as electrical distribution, ventilation and safety systems (typically adding $\sim$50\% to the cost). The overhead cost (estimated to be $\sim$22\% of the construction cost) includes design, development, management and inspection. Ultimately, the above favors compact and affordable technology. The integrated energy cost is the cost of power integrated over the time required to collect the data. This favors a high luminosity-per-wall-plug-power (i.e., both high energy efficiency and luminosity). The maintenance cost is the cost of paying personnel and replacement parts ($\sim$1\% of the construction cost per year), integrated over the full runtime of the programme, and is included to suppress the option of very low luminosity and power (but high luminosity-per-wall-plug-power) operating for a very long time. Finally, a carbon shadow cost is included (estimated at 800~EUR/ton \cite{EIB}), accounting for both construction and operating emissions---this is likely never actually paid in the form of a carbon tax, but is included to favor sustainability. 

In order to calculate the "Full Programme Cost" for a particular collider design, a detailed cost model has recently been developed, based largely on previous costings made by ILC and CLIC---this is an integral part of the new start-to-end simulation framework ABEL \cite{Chen2025}, built on its calculation of luminosity, length and power usage. The elements of the cost model, converted to ILC cost units (ILCUs; or 2012 dollars), are reported in Table 2 of Ref.~\cite{HALHF2025backup}.

\section{Bayesian optimization of collider parameters}

\begin{figure*}[t]
	\centering
    \includegraphics[width=0.9\textwidth]{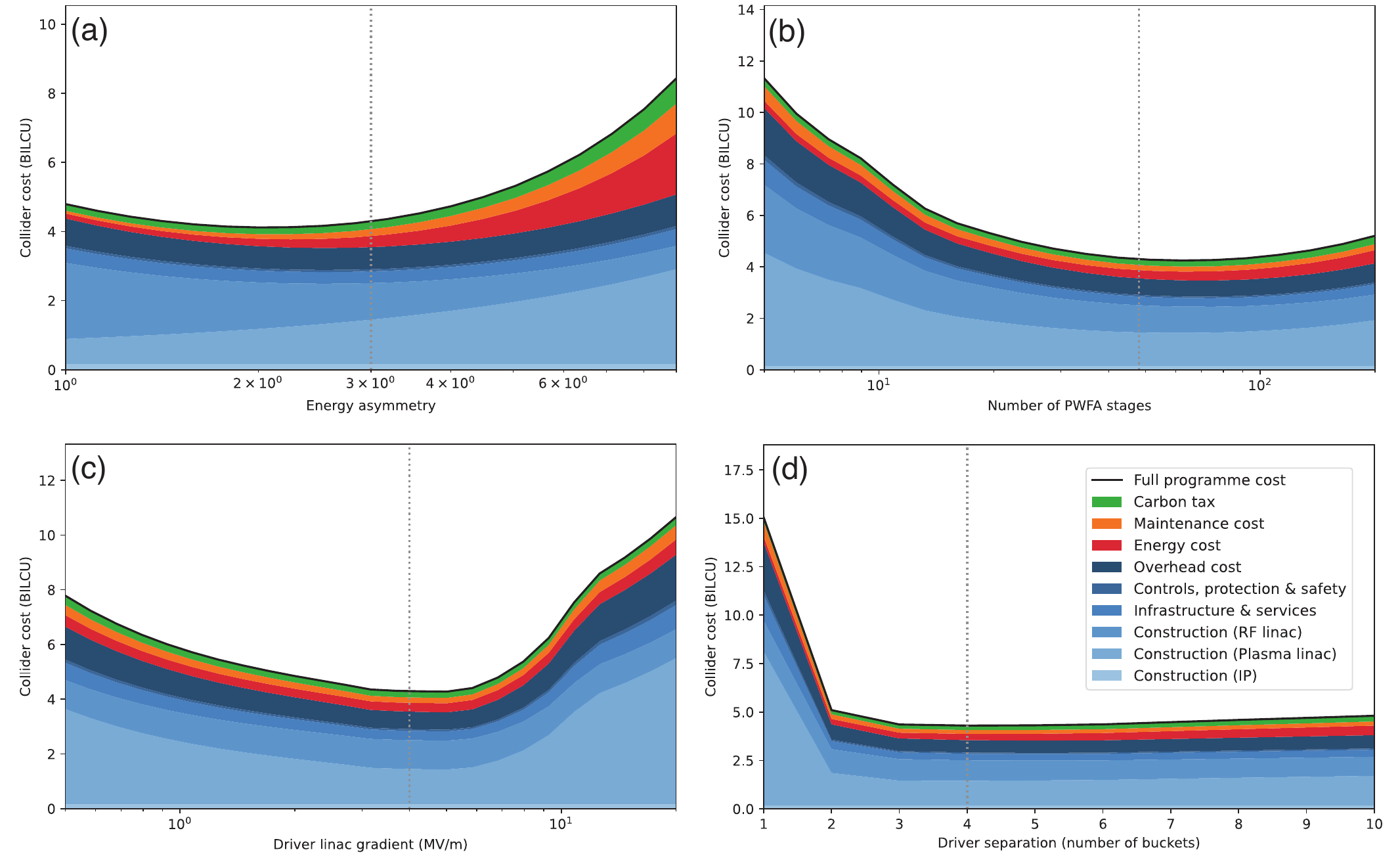}
\caption{Estimated collider cost as a function of four key parameters, keeping others constant: (a) the energy asymmetry, i.e.~the ratio of the electron energy to half the center-of-mass energy, (b) the number of PWFA stages, which also determines the driver energy and bunch-train length, (c) the driver RF linac gradient, for a given driver energy, and (d) the drive-bunch temporal separation, measured in RF buckets of \SI{1}{ns}. The updated baseline operating point is indicated with a dotted line.}
    \label{fig:parameter-variations}
\end{figure*}

In order to perform a global cost optimization of the collider, a set of 12 key parameters were identified: (1) the energy asymmetry; (2) the number of bunches in a train; (3) the repetition rate of the trains; (4) the combiner-ring compression factor; (5) the drive-bunch temporal separation; (6) the number of RF cells per structure in the driver linac; (7) the number of structures per klystron in the driver linac; (8) the accelerating gradient in the driver linac; (9) the accelerating gradient in the positron linac; (10) the number of RF cells per klystron in the positron linac; (11) the number of PWFA stages; and (12) the PWFA transformer ratio. As mentioned above, the plasma density and gradient were purposely not part of the optimization, as we wish to keep these as conservative and easily achievable as possible. The 12 parameters were simultaneously varied using Bayesian optimization; within ABEL this is implemented using the ``Ax'' framework \cite{Ax}. The optimization was run with 80 iterations, and converged to the same parameters every time, giving confidence that the global optimum had been found. 

Figure~\ref{fig:parameter-variations} showcases four different parameter variations. It can be seen that the updated baseline is close to (i.e., within ~10\%), but not identical to, the solution found by the Bayesian optimizer. This was the result of a conscious manual tuning of the parameters to cater to needs not well represented by the cost model: some were physical constraints, such as the need to have the number of stages be an integer multiple of the combiner-ring frequency multiplication factor (i.e., 12); others were more related to more aesthetic considerations, such as the desire to reduce overall length and power consumption if the added cost was small, e.g.~energy asymmetry, where the lowest overall cost was closer to 2 (i.e., \SI{250}{GeV} versus \SI{62.5}{GeV} for electrons and positrons, respectively) than the chosen value of 3---this produced the shortest length (\SI{5}{km}) and a lower construction cost.

Future work, toward a possible HALHF version 3, will involve performing the optimization with a higher level of fidelity in the start-to-end simulations (for more accurate estimates of luminosity and power usage), as well as the inclusion of more parameters. In particular, plasma density and gradient, should be part of the Bayesian optimization; this will require the inclusion of a detailed model of drive-beam misalignment and synchronization jitters, as well as plasma-density stability, in order to quantify the ideal balance between compactness and beam quality.

\section{Conclusion}

The new baseline design for HALHF, based on multidimensional Bayesian optimization and a detailed cost model, addresses many of the challenges identified in the original proposal. The HALHF Collaboration will continue to study and design the various subsystems, as well as push toward an even more cost-optimized design.

\section{ACKNOWLEDGEMENTS}
This work was supported by the the European Research Council (ERC Grant No. 101116161), the Research Council of Norway (NFR Grant No. 313770), the Leverhulme Trust (via an Emeritus Fellowship), DESY, the UK Science and Technology Facilities Council and the US Department of Energy. We acknowledge Sigma2 - the National Infrastructure for High-Performance Computing and Data Storage in Norway for awarding this project access to the LUMI supercomputer, owned by the EuroHPC Joint Undertaking, hosted by CSC (Finland) and the LUMI consortium. 

%
%
\ifboolexpr{bool{jacowbiblatex}}%
	{\printbibliography}%

\begin{thebibliography}{99} 

    \bibitem{ESPP2020}
    N. Mounet (ed.), 
    “European Strategy for Particle Physics - Accelerator R\&D Roadmap”, 
    CERN-2022-001 (CERN, 2022).

    \bibitem{P5}
    S. Asai \textit{et al.},
    “Pathways to Innovation and Discovery in Particle Physics - Particle Physics Project Prioritisation Panel Report”, 
    2023.

    \bibitem{CLIC}
    M. Aicheler \textit{et al.}, 
    ``A multi-TeV Linear Collider based on CLIC Technology : CLIC Conceptual Design Report", 
    CERN-2012-007 (CERN, 2012).

    \bibitem{ILC}
    T. Behnke (ed.) \textit{et al.}, 
    ``The International Linear Collider Technical Design Report", 
    (International Linear Collider, 2013).

    \bibitem{FCC}
    A. Abada et al., 
    ``FCC-ee: The Lepton Collider", 
    \textit{Eur. Phys. J. Spec. Top.} \textbf{228}, 261 (2019).
    
    \bibitem{Veksler1956}
    V. I. Veksler, 
    ``Coherent principle of acceleration of charged particles", 
    in Proc. CERN Symposium on High Energy Accelerators and Pion Physics (CERN, 1956), pp. 80--83.

    \bibitem{Fainberg1956}
    Ya. B. Fainberg, 
    ``The use of plasma waveguides as accelerating structures in linear accelerators", 
    in Proc. CERN Symposium on High Energy Accelerators and Pion Physics (CERN, 1956), pp. 84--90.
    
    \bibitem{Tajima1979}
    T. Tajima and J. M. Dawson,
    ``Laser electron accelerator",
    \textit{Phys. Rev. Lett.} \textbf{43}, 267 (1979).
    
    \bibitem{Chen1985}
    P. Chen, J. M. Dawson, R. W. Huff, and T. Katsouleas, 
    ``Acceleration of electrons by the interaction of a bunched electron beam with a plasma",
    \textit{Phys. Rev. Lett.} \textbf{54}, 693 (1985).
    
    \bibitem{Ruth1985}
    R. D. Ruth, A. W. Chao, P. L. Morton, and P. B. Wilson,
    ``A plasma wake field accelerator",
    \textit{Part. Accel.} \textbf{17}, 171 (1985).
    
    \bibitem{Esarey2009}
    E. Esarey, C. B. Schroeder and W. Leemans, 
    “Physics of laser-driven plasma-based electron accelerators”, 
    \textit{Rev. Mod. Phys.} \textbf{81}, 1229 (2009).

    \bibitem{LindstromCorde2025}
    C. A. Lindstr{\o}m, S. Corde \textit{et al.},
    “Beam-driven plasma wakefield acceleration”, 
    preprint at arXiv:2504.05558 (2025).
    
    \bibitem{Seryi2009}
    A. Seryi \textit{et al.},
   ``A Concept of Plasma Wake Field Acceleration Linear Collider (PWFA-LC)",
   in \emph{Proc. PAC’09}, Vancouver, Canada, May 2009, paper WE6PFP081, pp. 2688--2690.

    \bibitem{Adli2013}
    E. Adli \textit{et al.},
    ``A Beam Driven Plasma-Wakefield Linear Collider: From Higgs Factory to Multi-TeV"
    in \emph{Proc. of the 2013 US Community Study on the Future of Particle Physics}, 2013.
    
    \bibitem{Foster2023}
    B. Foster, R. D'Arcy and C. A. Lindstr{\o}m, 
    “A hybrid, asymmetric, linear Higgs factory based on plasma-wakefield and radio-frequency acceleration”, 
    \textit{New J. Phys.} \textbf{25}, 093037 (2023).

    \bibitem{Cao2024}
    G. J. Cao \textit{et al.},
    ``Positron acceleration in plasma wakefields",
    \textit{Phys. Rev. Accel. Beams} \textbf{27}, 034801 (2024).

    \bibitem{Lindstrom2021a}
    C. A. Lindstr{\o}m, 
    “Staging of plasma-wakefield accelerators”, 
    \textit{Phys. Rev. Accel. Beams} \textbf{24}, 014801 (2021).

    \bibitem{Foster2025}
    B. Foster \textit{et al.},
    “Proceedings of the Erice workshop: A new baseline for the hybrid, asymmetric, linear Higgs factory HALHF”, 
    \textit{Phys. Open} \textbf{23}, 100261 (2025).

    \bibitem{HALHF2025}
    E. Adli \textit{et al.} (HALHF Collaboration),
    “HALHF: a hybrid, asymmetric, linear Higgs factory using plasma- and RF-based acceleration”, 
    arXiv:2503.19880 (2025).
    
    \bibitem{HALHF2025backup}
    E. Adli \textit{et al.} (HALHF Collaboration),
    “HALHF: a hybrid, asymmetric, linear Higgs factory using plasma- and RF-based acceleration. Backup Document”, 
    arXiv:2503.23489 (2025).
    
    \bibitem{Vernieri2023}
    C. Vernieri \textit{et al.},
    “A “Cool” route to the Higgs boson and beyond. The Cool Copper Collider”, 
    \textit{J. Instrum.} \textbf{18}, P07053 (2023).

    \bibitem{Baynham2009}
    E. Baynham \textit{et al.},
    “The development of a superconducting undulator for the ILC positron source”, 
    in \textit{Proc. PAC’09}, Vancouver, Canada, May 2009, paper WE2RAI01, pp. 1839--1843. 

    \bibitem{Cilento2021}
    V. Cilento \textit{et al.},
    ``Dual beam delivery system serving two interaction regions for the compact linear collider"
    \textit{Phys. Rev. Accel. Beams} \textbf{24}, 071001 (2021).
    
    \bibitem{EIB}
    EIB Group Climate Bank Roadmap 2021–2025, 
    \url{doi:10.2867/503343}
    
    \bibitem{Chen2025}
    J. B. B. Chen \textit{et al.},
    ``ABEL: The adaptable beginning-to-end linac simulation framework", 
    presented at the 16th International Particle Accelerator Conf. (IPAC’25), Taipei, Taiwan, June 2025 paper TUPS012, this conference.
    
    \bibitem{Ax}
        Adaptive Experimentation Platform “Ax”, \url{ax.dev}.

	\end{thebibliography}
	{%
	
	
} 
%
%


\end{document}